\documentclass[pre,aps,floats,superscriptaddress,preprint]{revtex4-1}
\usepackage{amsmath,amssymb}
\usepackage{graphicx}
\usepackage{psfrag}

\def\beq{\begin{equation}}
\def\eeq{\end{equation}}
\def\bea{\begin{eqnarray}}
\def\eea{\end{eqnarray}}
\begin{document}
\title{Continuous Universality in  non-equilibrium relaxational dynamics of
$O(2)$ symmetric systems}
\author{Niladri Sarkar}\email{niladri.sarkar@saha.ac.in}
\author{Abhik Basu}\email{abhik.basu@saha.ac.in}
\affiliation{Theoretical Condensed Matter Physics Division, Saha
Institute of Nuclear Physics, Calcutta 700064, India}



\date{\today}

\begin{abstract}
 We elucidate a non-conserved relaxational nonequilibrium dynamics of a $O(2)$
 symmetric model. We drive the system out of equilibrium by
 introducing a non-zero noise cross-correlation of amplitude $D_\times$ in a stochastic
 Langevin description of the system, while maintaining the $O(2)$
 symmetry of the order parameter space. By performing dynamic
 renormalization group calculations in a field-theoretic set up, we analyze the ensuing nonequilibrium
 steady states and evaluate the scaling
 exponents near the critical point, which now depend explicitly on
 $D_\times$. Since the latter
 remains unrenormalized, we obtain universality classes varying
 continuously with $D_\times$. More interestingly, by changing
 $D_\times$ continuously from zero, we can make our system move away
 from its equilibrium behavior (i.e., when $D_\times=0$) continuously and
 incrementally.
\end{abstract}

 \maketitle

\section{Introduction}

The concept of universality near  critical points in equilibrium
systems has a long history and is theoretically well-developed \cite{fisherrev}.
When equilibrium systems undergo second order phase transition at a
critical point, they display universal scaling properties for
thermodynamic quantities and correlation functions. These are
characterized by a set of scaling exponents, which are {\em
universal} in the sense that they depend only on the spatial
dimension $d$ and the symmetry of the order parameter (e.g., Ising,
XY etc.) \cite{fisherrev}, but not on the parameters that specify
the (bare) Hamiltonian. Notable exceptions are the $2d\;XY$ model
and the related models, where the renormalization group flow is
characterized by a fixed line and consequently the scaling exponents
exhibit a continuous dependence on the value of the bare stiffness
parameter that appears in the model Hamiltonian. The idea of
universality may be readily extended to equilibrium dynamics close
to critical points, where the systems  exhibit universality through
the dynamic scaling exponents, which characterize the time-dependent
unequal-time correlation functions. Their universality classes
depend upon the presence or absence of conservation laws and the
non-dissipative (reactive) terms in the underlying dynamical
equations \cite{halp}. For driven dissipative out of equilibrium
systems with nonequilibrium steady states (NESS), the general
picture about universality is still wide open. In the recent past,
attempts with significant success have been made in classifying the
physics of non-equilibrium systems at long time and large length
scales into universality classes. For example,  the robustness of
the standard universality classes in critical dynamics to
detailed-balance violating perturbations are shown in
Ref.~\cite{tauber-etal:02}. In addition only models having conserved
order parameters and spatially anisotropic noise correlations
exhibit novel features. In contrast, recent works demonstrate that
truly non-equilibrium dynamic phenomena, whose steady states cannot
be described in terms of Gibbsian distributions, are rather
sensitive to all kinds of perturbations. Well-known examples include
driven diffusive models~\cite{zia:review}, and fluid- and
magnetohydrodynamic-
turbulence~\cite{biskampbook,rahulrev,abthesis}. Overall, in
contrast to equilibrium systems, how one may classify the
universality classes for systems out of equilibrium remains an
unresolved issue till the date. It is well-known that for models
driven out of equilibrium, not only dynamical properties but even
the static properties (e.g., the static correlation functions)
depend crucially on the distributions of noises which appear in a
Langevin description of the model. In light of this a  useful
strategy is to investigate nonequilibrium universality classes is to
construct simple models with non-thermal noises, whose dynamics will
reveal this sensitive dependence of universal properties on noise
distributions in a systematic manner.

In this article we examine the particular issue of universality in
non-equilibrium for a simple relaxational dynamics (model A in the
terminology of Ref.~\cite{halp}) of an $O(2)$-symmetric system
(equivalently, the {\em classical XY model}) in dimension
$d=d_c-\epsilon$ where the upper critical dimension of the model
$d_c=4$. The $O(2)$ symmetric model is a special case with $N=2$ for
the more general $O(N)$ model. The equilibrium critical dynamics of
these models are discussed in details in Refs.~\cite{halp}. The
model is forced out of equilibrium by specific choices of the
variances of the additive noises in the Langevin equations for the
dynamical variables (see below). Phase transitions and associated
universal properties at the critical point in systems with
relaxational (model- A type in the language of Ref.~\cite{halp})
dynamics have been shown to be remarkably robust against the
introduction of various competing dynamics which are local and do
not conserve the order parameter~\cite{grin}, including those which
breaks the discrete symmetry of the system~\cite{bass}. We show that
its NESS depend sensitively on the parameters of the model. We use
field theoretic renormalization group calculation
\cite{medina,amit,epsilon1,bausch,veltman} using dimensional
regularization \cite{veltman} based on an $\epsilon$-expansion
\cite{amit} scheme.

Our principal results are: (i) As a temperature-like variable in the
model (see below) is lowered, our model undergoes a phase transition
from a {\em high temperature} paramagnetic disordered phase to a
{\em low temperature} ferromagnetic ordered phase undergoing a
second order phase transition at a nonequilibrium critical point,
(ii) Universal scaling behavior near the critical point determined
by a set of standard scaling exponents characterizing the
correlation and the response functions that depend explicitly on the
magnitude of the noise cross-correlations; in effect we obtain a
continuous universality parameterized by the noise
cross-correlations, the latter being a marginal operator in the
model. The remainder of the paper is organized as follows: In
Sec.~\ref{model} we set up our continuum $O(2)$ symmetric dynamical
model for a non-conserved order parameter to study its universal
properties near the critical point. We introduce noises which break
the Fluctuation-Dissipation-Theorem (FDT) \cite{fdt}, and thus drive
the system out of equilibrium, but keep the rotational invariance in
the order parameter space unbroken. In the next Sec.~\ref{results}
we set up the field theoretic formulation for our model in terms of
a path integral description. We use a diagrammatic perturbation
theory and calculate fluctuations corrections to different vertex
functions up to the two-loop order. We then use a minimal
subtraction scheme to calculate different critical exponents within
an $\epsilon$-expansion. In Sec.~\ref{conclu} we summarize and
discuss the implications of our results.

\section{Model equations}
\label{model}

In this Section we set up our model equations to describe a
simple nonequilibrium generalization of the relaxational (model A)
dynamics in the overdamped limit of a non-conserved $O(2)$ symmetric
order parameter. The equilibrium characteristics of this dynamical
model have been extensively discussed in the literature, see, e.g.,
Ref.~\cite{halp}. We consider a second order phase transition
described by a vector order parameter $\phi_i,\,i=1,\,2$. As
we furthermore assume isotropy in order parameter space, the static
critical properties are described by an $O(2)$-symmetric
$\phi^4$-type Landau-Ginzburg-Wilson free energy functional in $d$
space dimensions,
\begin{equation}
F[\phi_i]=\int d^dx [\frac{\tau}{2} ({\phi_1}^2 +{\phi_2}^2) +
\frac{1}{2}\{(\nabla \phi_1)^2 + (\nabla\phi_2)^2\} + \frac{u}{4!}
({\phi_1}^2 +{\phi_2}^2)^2], \label{free}
\end{equation}
where $\tau = (T - T_{c} )/T_{c}$ is the bare relative distance from
the mean-field critical temperature $T_{c}$ and $u>0$ is a (bare)
coupling constant. The free energy functional $F$ is manifestly
rotation invariant in the order parameter space. This $F$ determines
the equilibrium probability distribution for $\phi_i$. Free energy
functional $F$ allows us to compute any of the two (independent)
critical exponents, e.g., the anomalous dimension $\eta$ and the
correlation length exponent $\nu$, by means of renormalization group
procedure, based on diagrammatic perturbation theory with respect to
the static non-linear coupling $u$ within a systematic expansion in
terms of $\epsilon = 4 - d$ about the static upper critical
dimension $d_c = 4$. These exponents have well-defined physical
meaning. For example, the exponent $\eta$ characterizes how the
order parameter correlation function at criticality decays in a
spatial power-law fashion, $\langle \phi_i ({\bf r})\phi_j ({\bf
r'})\rangle \propto 1/|{\bf r-r'}|^{d-2+\eta}\delta_{ij}$, or
equivalently of the static susceptibility $\chi ({\bf q})\propto
1/q^{2-\eta}$ where $\bf q$ is a wavevector, and the exponent $\nu$
describes how the correlation length $\xi$ diverges as the
renormalized critical temperature $T_c$ is approached, $\xi\propto
|T -T_c|^{-\nu}$. Further, the fluctuation-corrected true transition
temperature $T_c$ is smaller as compared to the mean-field critical
temperature $T_{0c}$ , i.e., $\tau_{0c} = T_c - T_{0c} < 0$.

In contrast to equilibrium systems, for systems out of equilibrium,
there is no detailed balance and even the static quantities must be
calculated from the underlying dynamics directly. The
description of the dynamics of such systems in terms of continuum
degrees of freedom are often based on stochastically driven Langevin
equations of motion for the relevant dynamical degrees of freedom.
For Langevin equations describing processes relaxing towards a
thermal equilibrium state the correlation functions of a given
degree of freedom and the corresponding susceptibility are connected
through the FDT which in turn fixes specific relations between the
variances of the noises and the diffusivities. For example, the
non-conserved relaxational (model A) dynamics for a vector order
parameter $\phi_i$ is given by
\begin{equation}
 {\partial\phi_i \over \partial
t}=-\Gamma{\delta F \over \delta\phi_i} + g_i,
\end{equation}
 where $i=1,2$; $F$
is given by (\ref{free}), $\Gamma$ is a kinetic coefficient and
$g_i$ are temporally delta-correlated zero-mean Gaussian
stochastic noises with specified variances. Assuming spatial
translational invariance we can write generally
\begin{equation}
\langle g_i ({\bf q},t)g_j ({\bf -q},0)\rangle = 2D_{ij}({\bf
q})\delta (t).
\end{equation}
If we now set $D_{ij}({\bf k},t)=2K_BT \Gamma\delta_{ij}$, where
$K_B$ is the Boltzmann constant and $T$ is the temperature, then the
FDT is obeyed and the corresponding Fokker-Planck equation admits a
steady-state equilibrium solution $P_{eq}\sim \exp [-F/K_BT]$. In
contrast, in nonequilibrium situations there are no general
relations linking the noise variance and the kinetic coefficients
and the FDT is broken. Since noises in a Langevin description
describe the effects of the environment (e.g., thermal baths), such
nonequilibrium noises reflect external drives. What are the simplest
choices of the noise variances which explicitly break the FDT,
without having to break the $O(2)$ symmetry? One possible way to do
that is to introduce two different noise strengths in the noise
correlation matrix and break the FDT. This can be realized by the
choice
 \begin{eqnarray}
\langle g_1({\bf q},t)g_1(-{\bf q},0)\rangle &=& 2 \Gamma D_1 \delta(t), \nonumber \\
\langle g_2({\bf q},t)g_2(-{\bf q},0)\rangle &=& 2 \Gamma D_2 \delta(t), \nonumber \\
\langle g_1({\bf q},t)g_2(-{\bf q},0)\rangle &=& 0.
\end{eqnarray}
Such a choice as above will certainly break the FDT but
unfortunately will break the $O(2)$ symmetry of the ensuing dynamics
as well. Ref.~\cite{racz} investigated nonequilibrium critical
properties of O(n)-symmetric models with reversible mode-coupling
terms. Specifically, a variant of the model of Sasv\'ari, Schwabl,
and Sz\'epfalusy (SSS) is studied, where violation of detailed
balance is incorporated by allowing the order parameter and the
dynamically coupled conserved quantities to be governed by heat
baths of different temperatures. They however find that upon
approaching the critical point detailed balance is restored, and the
equilibrium static and dynamic critical properties are recovered.
Yet another option is to couple the system with the corresponding
conserved angular momentum and introduce dynamical anisotropy in the
noise for the conserved quantities, i.e., by constraining their
diffusive dynamics to be at different temperatures $T_\parallel$ and
$T_\perp$ in $d_\parallel$- and $d_\perp$-dimensional subspaces,
respectively, see Ref.~\cite{uwe1} for detailed calculation for the
SSS model for planar ferro- and isotropic antiferromagnets.
Ref.~\cite{uwe1} showed that the equilibrium fixed point (with
isotropic noise) to be stable with respect to these non-equilibrium
perturbations, and the familiar equilibrium exponents therefore
describe the asymptotic static and dynamic critical behavior. Novel
critical features are only found in extreme limits, where the ratio
of the effective noise temperatures $T_\parallel/T_\perp$ is either
zero or infinite. In a similar study, Ref.~\cite{uwe11}
discussed nonequlibrium dynamics in a liquid-gas model with
reversible mode couplings. The model is driven out of equilibrium by
introducing different temperatures for different dynamical
variables, or, by having anisotropic noises. However, no new genuine
nonequilibrium stable fixed point is found (within one-loop
calculations). Similar approaches to nonequilibrium critical
dynamics of the relaxational models C and D (in the language of
Ref.~\cite{halp}) are discussed in Ref.~\cite{uwe2} and involve
coupling a non-conserved and conserved order parameter,
respectively, to a conserved density, with the order parameter and
density fields are being in contact with heat baths at different
temperatures. Within a one-loop calculation it finds, in certain
cases, {\em continuously varying} static and dynamic critical
exponents, as a function of a dimensionless nonequilibrium parameter
in the model. An alternative route to introduce detailed
balance violation in the simple model A-type relaxational dynamics 
for the order parameter $\phi_i$is to introduce non-zero noise cross correlations
which will make the noise matrix off-diagonal. This will break the
FDT as the noise matrix is then not proportional to the kinetic
coefficient matrix (which is proportional to the unit matrix in the
present case). We take cross noise strengths as $\hat{D}({\bf q})$.
We write
\begin{eqnarray}
\langle g_1({\bf q},t)g_1(-{\bf q},0)\rangle&=&\langle g_2({\bf q},t)g_2(-{\bf q},0)\rangle=2 D\Gamma \delta(t) \nonumber \\
\langle g_1({\bf q},t)g_2(-{\bf q},0)\rangle &=& 2\hat{D}({\bf
q})\Gamma \delta(t). \label{noisevar}
\end{eqnarray}
In general the function $\hat D({\bf q})$ is a complex function of
wavevector $\bf q$.

The form of the function $\hat{D}({\bf q})$ may be restricted by
demanding rotational invariance of the noise variance matrix
(equivalently by demanding $O(2)$ symmetry of the dynamics). Under a
rotation by an arbitrary angle $\theta$ in the order parameter space
the noise variance matrix transforms to \bea N'=\Gamma\left(
\begin{array}{cc}
\cos\theta & \sin\theta \\
-\sin\theta & \cos\theta
\end{array}\right)\left(
\begin{array}{cc}
D & \hat{D} \\
\hat{D}^* & D
\end{array}
\right)\left(
\begin{array}{cc}
\cos\theta & -\sin\theta \\
\sin\theta & \cos\theta
\end{array}\right)
.\label{noisemat} \eea where the noise variance matrix before
rotation is \bea N=\Gamma\left(\begin{array}{cc}
D & \hat{D} \\
\hat{D}^* & D
\end{array}
\right). \eea
 Now we demand $N=N'$ due to rotational invariance. This after a simple algebra then yields that the noise cross correlation amplitude should be fully
imaginary or $\hat{D}({\bf q})=-\hat{D}^*({\bf q})$. Since in the
real space $\hat D({\bf r})$ must be a real function, we find $\hat
D({\bf q})$ must be an odd function of $\bf q$. In order for the
noise cross correlation to have the same na\"ive dimension as $D$
(so that both $D$ and $\hat D$ are equally {\em relevant} in an RG
sense), we set $\hat D({\bf q})\hat D({\bf q})=D_\times ^2$ where
$D_\times ^2$ is a constant (and has the same dimension as $D^2$).
We henceforth replace $\hat{D}({\bf q})$ by $i\hat D({\bf q})$ where
$\hat D({\bf q})$ is now completely real. This reflects the fully
imaginary nature of the cross correlation explicitly. Thus the
explicit forms of the two equations of motion for $\phi_1$ and
$\phi_2$ are
\begin{eqnarray}
{1 \over \Gamma}{\partial\phi_1 \over \partial t} &=& -\tau\phi_1 + c\nabla^2\phi_1 - {u \over 3!}\phi_1^3 - {u \over 3!}\phi_1\phi_2^2 + {g_1 \over \Gamma}, \nonumber \\
{1 \over \Gamma}{\partial\phi_2 \over \partial t} &=& -\tau\phi_2 +
c\nabla^2\phi_2 - {u \over 3!}\phi_2^3 - {u \over 3!}\phi_2\phi_1^2
+ {g_2 \over \Gamma} , \label{basiceom}
\end{eqnarray}
complemented by the noise variances as below:
\begin{eqnarray}
\langle g_1({\bf q},t)g_1(-{\bf q},0)\rangle&=&\langle g_2({\bf q},t)g_2(-{\bf q},0)\rangle=2 D\Gamma \delta(t) \nonumber \\
\langle g_1({\bf q},t)g_2(-{\bf q},0)\rangle &=& 2i\hat{D}({\bf
q})\Gamma \delta(t).\label{finalnoise}
\end{eqnarray}
One may in addition consider including a conserved angular momentum
as a slow variable in the problem (see, e.g., model E in
Ref.~\cite{halp}). We do not do that here for simplicity. Model
equations (\ref{basiceom}) suffices for our purposes of exploring
nonuniversal features in a simple setting. Are there limits on the
value of $D_\times$ in this model? To obtain that we demand the
noise variance matrix to have eigenvalues which are real positive or
zero. The eigenvalues concerned are $D\pm D_\times$. Thus
$|D_\times| \leq D$, or in terms of a dimensionless number
$N_\times=(D_\times/D)^2$, $N_\times\leq 1$. In the subsequent
calculations we will find that $N_\times$ enters into the
expressions of different scaling exponents explicitly.

 Equations of motion
(\ref{basiceom}) are written in an $O(2)$ invariant representation.
Using the equivalence between $O(2)$ and $U(1)$ representations, one
may write an equivalent $U(1)$ representation of the dynamics. The
free energy in the $U(1)$ representation takes the form
\begin{equation}
F_U[\psi\psi^*]=\int d^dx [\tau\psi\psi^* + (\nabla \psi)
(\nabla\psi^*) + \frac{u}{3!} (\psi\psi^*)^2], \label{u1}
\end{equation}
where complex fields $\psi={1 \over \sqrt{2}}(\phi_1+i\phi_2)$; $\psi^*$ is the
complex conjugate of $\psi$. The corresponding Langevin equations
of motion in the overdamped limit are given by
\begin{equation}
\frac{\partial\psi}{\partial t}=-\Gamma \frac{\delta
F_U}{\delta\psi^*} +\xi,
\end{equation}
where zero-mean Gaussian distributed complex noise $\xi$ has the
following correlations in the Fourier space:
\begin{eqnarray}
\langle\xi ({\bf q},t)\xi({\bf -q},0)\rangle=0=\langle \xi^*({\bf
q},t)\xi^*({\bf -q},0\rangle,\nonumber \\
\langle \xi({\bf q},t)\xi^*(-{\bf q},0)\rangle = 2D\Gamma \delta (t)
+2i\hat D ({\bf q}) \Gamma\delta (t).
\end{eqnarray}
Thus introduction of noise cross-correlations in the $O(2)$
description is equivalent to adding an imaginary and odd function of
$\bf q$ in the variance $\langle\xi \xi^*\rangle$. Before we embark
upon detailed calculation let us consider possible physical
(microscopic) realizations of our continuum model in terms of 
stochastic lattice-gas models. However, what we discuss below
does not fully and precisely define a microscopic model, but rather
outlines broad features that an eventual appropriate microscopic
realization should posses. Consider a system of XY ($(O(2)$ spins)
either on a (hypercubic) lattice or a continuum in $d$-dimensions,
interacting with an additional mobile species in the system which
diffuses randomly, undergoing symmetric exclusion process (SEP) to any
of the nearest sites, if vacant. A simple model of interaction
between these two species could be where each diffusing particle
carries an XY spin attached to it, and the nearest-neighbor exchange
coupling $J_{ij}$ that defines the XY model is related to the local
particle density $n_i(t)$ at site $i$ via $J_{ij}\propto n_i (t)
n_j(t)$. Next, noting that the microscopic dynamics of both
the spins and particles are stochastic, characterized by two sets of
random numbers $\tilde g_{1i}(t)$ and $\tilde g_{2i}(t)$,
respectively, we impose that $\tilde g_1$ and $\tilde g_2$ are
cross-correlated, with the cross-correlation function being of the
form (in the continuum limit) $A\delta ({\bf x}_1-{\bf x}_2) +
B({\bf x}_1 - {\bf x}_2)$, where ${\bf x}_1$ and ${\bf x}_2$
are two points in the lattice, $A$ is a numerical constant and
$B({\bf x})$ is an odd function of position $\bf x$ having the same
dimension as $\delta ({\bf x}_1 - {\bf x}_2)$. The presence of the
odd function $B({\bf x}_1 - {\bf x}_2)$ in the cross-correlation
function ensures lack of reflection invariance of the underlying
stochastic microscopic dynamics. Thus the measured quantities (e.g.,
correlation functions of appropriate densities) should reflect this
lack of reflection invariance. At this level, the dynamics for the
additional species is clearly conserving. This implies there will be
no timescale present on which the additional variables of the mobile
species can be treated as fast and eliminated to yield an effective
equation of motion for the XY spins alone with the effects of the
diffusing species buried in the additive noises in the effective
spin equations. However, if particle nonconservation is introduced,
e.g., via evaporation-deposition affects, the local particle density
dynamics will be fast and then may be eliminated to produce an
effective spin dynamics. Since the noises in such effective theories
contain information about the already eliminated fast degrees of
freedom (in this case the local diffusing particle density), there
will be non-zero noise cross-correlations of specified structures as
above, due to the particular chosen structure of the underlying
reflection invariance breaking microscopic dynamics. Alternatively,
one may introduce the driving as a temporally delta-correlated
fluctuating magnetic field ${\bf h}({\bf x},t)=(h_x,h_y)$ with $h_x$
and $h_y$ having short ranged spatial correlations as in
(\ref{finalnoise}). In both the cases, noise cross-correlations of
appropriate structures will be generated in the effective Langevin
description. With this short background in mind, let us now
investigate the universal scaling properties of the model described
by the Langevin equations (\ref{basiceom}) together with the noise
variances (\ref{finalnoise}).
 The presence of non-linear terms in Eqs.~(\ref{basiceom})
rules out exact solutions, and we resort to perturbative
calculations  that we discuss below.

\section{Nonequilibrium steady states}
\label{results}

Let us first consider the high temperature phase of the system. At
high temperature with $\tau>0$, the system is in the paramagnetic
phase, i.e., $\langle \phi_i ({\bf x},t\rangle =0$, where
$\langle...\rangle$ means averaging over the noise distributions.
The correlation length $\xi$ remains finite for all $\tau>0$. The
only effect of the noise cross-correlations is to make the
cross-correlation function $\langle\phi_1 ({\bf
x},t)\phi_2(0,0)\rangle$ non-zero with a finite correlation length
$\xi$. Further, as in equilibrium critical dynamics, the
paramagnetic phase is linearly stable and the fluctuations have a
finite life time for all wavevector. Nevertheless, the FDT is
violated for all $\tau>0$ due to the noise cross-correlations.

Near the critical point, the system becomes scale invariant and the
correlation length $\xi$ diverges, leading to the emerging
macroscopic physics near the critical point being vastly different
from the paramagnetic phase. A quantitative description of the
nature of correlations near the critical point requires the
principles and formalisms of the Dynamic Renormalization Group
(DRG), which we execute here by using a field-theoretic framework.
Detail discussions of the technical aspect of field-theoretic DRG
calculations are well-documented in the literature, see, e.g.,
Refs.~\cite{medina,uwe-book}. In order to set up the background let
us examine the linearized version of the model equations
(\ref{basiceom}) together with the noise correlations
(\ref{noisevar}) at $\tau=0$ by dropping all the non-linear terms
($u=0$). The system remains $O(2)$ invariant, but the FDT is already
broken at this linear level due to the noise cross-correlations.
Obviously, the field correlations from the linearized model
equations can be exactly calculated. In this linear theory, in the
critical region, defined by $\tau=0$, the linear theory is massless
resulting in divergent long wavelength fluctuations, as can be seen
by explicit calculations of the correlation functions
$C_{ij}=\langle \phi_i ({\bf x},t)\phi_j (0,0)\rangle,\,i=1,2$,
which may be written down in a scaling form at the critical point
$\tau=0$:
\begin{equation}
C_{11}({\bf x},t)=C_{22}({\bf x},t)=
x^{2-d}f(t/x^z),\label{scalingform}
\end{equation}
where $f$ is an analytic function of its argument. For the
cross-correlation function $C_{12}$ (and by symmetry $C_{21}$)
displays the same scaling form, but with a different amplitude and
is an odd function of $\bf x$.

 What is the nature of these diverging fluctuations when the non-linear
 terms are present ($u>0$)? The presence of nonlinear terms no longer allows for exact solutions, in contrast
 to the linearized theory. However, this question may be systematically addressed via
 standard implementation of DRG procedure, based on a perturbative expansion
 in the {\em small} coupling $u$ about the linear theory. The perturbative
 corrections to the correlation
function may be equivalently viewed as arising from modifications
(renormalization) of the parameters $\tau,\,u,\,\Gamma,\,D$ and the
dynamical fields $\phi_1$ and $\phi_2$. Renormalizability of the
theory ensures that correlator $C_{ij}$ will retain scaling forms
similar to (\ref{scalingform}) with exponents different from those
appearing in (\ref{scalingform}) and new scaling functions at
(renormalized) $\tau=0$:
\begin{equation}
C_{11}({\bf x},t)=C_{22}({\bf
x},t)=x^{2-d-\eta}f_s(t/x^z),\label{newscaling}
\end{equation}
where $\eta$ and $z$ are the anomalous dimension and dynamic
exponents respectively, and $f_s$ is a new scaling function
\cite{dynamic}. For the linear theory described above, $\eta=0$ and
$z=2$. The nonlinear coupling $u$ is expected to change these
exponents for the linear theory. The equilibrium critical dynamics
of several nonlinear problems have been described in
Ref.~\cite{halp}. In our subsequent analysis below we assume
renormalizability and justify it {\em post facto} by a low order (up
to two-loop) perturbation theory.

Operationally, the DRG procedure is conveniently performed in terms
of a path integral description based on a dynamic generating
functional which is to be constructed out of the Langevin equations
(\ref{basiceom}) and the corresponding noise variances given by
Eq.~(\ref{finalnoise}) following standard procedures
\cite{bausch,genfunc}. The dynamic generating functional for the
present model is given by
 \bea
{\mathcal Z} &=& \int {\mathcal D}\phi_1 {\mathcal D}\phi_2
{\mathcal D}\hat{\phi}_1 {\mathcal D}\hat{\phi}_2 \exp[-{D \over
\Gamma}\int \frac{d^dk}{(2\pi)^d} \hat{\phi}_1\hat{\phi}_1 - {D
\over \Gamma}
\int \frac{d^dk}{(2\pi)^d} \hat{\phi}_2\hat{\phi}_2 - i\int \frac{d^dk}{(2\pi)^d} \hat{\phi}_1{\hat{D}_k \over \Gamma}\hat{\phi}_2] \nonumber \\
&\times& \exp[i\int \frac{d^dk}{(2\pi)^d} \hat{\phi}_1 \{{1 \over
\Gamma}\partial_t \phi_1 + {\delta F \over \delta \phi_1} \}]
\exp[i\int \frac{d^dk}{(2\pi)^d} \hat{\phi}_2 \{{1 \over \Gamma}\partial_t \phi_2 +
{\delta F \over \delta \phi_2} \}],\label{gen}
 \eea
 where $\hat{\phi}_1$ and $\hat{\phi}_2$ are conjugate response
fields corresponding  to the order parameter fields $\phi_1$ and
$\phi_2$ used to obtain the noise-averaged generating functional
$\mathcal Z$. Clearly, with our choice of the noise
cross-correlations, the generating functional respects the $O(2)$
symmetry of the underlying dynamics, but explicitly breaks the FDT.
It now remains to be seen whether this breakdown of the FDT remains
valid even for the {\em effective, renormalized} version of this
theory, or it is restored in the long wavelength limit. In the above
we have used the Ito prescription \cite{janssen1} while writing down
the generating functional (\ref{gen}).

The perturbative calculational framework begins with the
construction of the perturbation expansion of one-particle
irreducible Feynman diagrams for all possible correlation and the
corresponding vertex functions constructed out of the fields
$\phi_1$, $\hat{\phi}_1$, $\phi_2$ and $\hat{\phi}_2$. Such a
perturbation expansion is meaningful only wheen the coupling $u$ is
small. This is generally accomplished using the standarad DRG
procedure which involves an order by order expansion about $d_c-d$,
where $d_c$ is the upper critical dimension and $d$ the physical
space dimension. As a result, the renormalized coupling $u_R$ flows
to small values, of order $\epsilon=d_c-d$. As for equilibrium field
theories, a straight forward scaling analysis yields that $d_c=4$
for this model: We scale $x\rightarrow b_xx$ and $t\rightarrow b_tt$
in the action, where $b_x>1$ and $b_t>1$ are arbitrary parameters,
and find out how the other quantities scale to maintain scale
invariance. It is seen from the bare equation leaving aside the
non-linear terms we must have $b_t=b_x^2=b^2$ in order to have
dynamical scaling, where we have taken $b_x=b$. For the action to
remain invariant, the fields $\phi_1$ and $\phi_2$ pick up a
canonical dimension $d/2-1$ and the coupling constant $u$ a
dimension $4-d$. Hence the critical dimension at which the coupling
constant $u$ becomes dimensionless is $d_c=4$. This yields that the
perturbation theory will be infra-red (IR)-singular for $d\leq 4$,
and consequently the system will show non-trivial critical behavior
in that regime, while for $d \geq 4$ the perturbation theory
contains ultra-violet (UV) divergences, and the (static) mean-field
exponents together with dynamic exponent $z=2$ (dynamic exponent of
the linearized theory) will describe the system at the critical
point. To make the field theory UV renormalized it is needed to
introduce the multiplicative renormalization constants in order to
render all the non-vanishing two- and four-point functions finite.
Within the DRG procedure this is achieved by demanding the
renormalized vertex functions in the theory, or their appropriate
momentum and frequency derivatives, to be finite when the
corresponding loop-integrals representing fluctuation corrections
are considered as functions of conveniently chosen frequency and
momentum, well outside the IR regime. In order to compute the
associated one- and two-loop momentum integrals we employ the
dimensional regularization scheme and choose $\tau=\mu^2$ as our
normalization point, where $\mu$ is an intrinsic momentum scale of
the renormalized theory. From the renormalization constants
($Z$-factors) that render the underlying field theory finite in the
ultraviolet (UV), one may then derive the RG flow equation, which
describes how correlation functions change under scale
transformations. Since the theory becomes scale-invariant in the
vicinity of a critical point (or an RG fixed point),  one may employ
the information previously gained about the UV behavior to access
the physically interesting power laws governing the infrared (IR)
regime at the critical point ( $\tau \propto T-T_c\rightarrow 0$)
for long wavelengths (wavevector ${\bf q}\rightarrow 0$) and low
frequencies ($\omega\rightarrow 0$). The scaling behavior of the
correlation or vertex functions may be extracted by finding their
dependence on $\mu$ by using the RG equation.

Ward identities due to the rotational invariance of the model (in
the order parameter space) ensures the following exact relations
between different vertex functions: \bea
 \Gamma_{\hat\phi_1\phi_1} ({\bf k},
\omega) &=& \Gamma_{\hat\phi_2\phi_2}({\bf k},\omega)  = \Gamma_{11}({\bf
k},\omega), \\
 \Gamma_{\hat\phi_1\hat\phi_1}({\bf
k},\omega) &=& \Gamma_{\hat\phi_2\hat\phi_2}({\bf
k},\omega)=\Gamma_{20}({\bf k},\omega)
\eea
and
\bea
&&\Gamma_{\hat\phi_1\phi_1\phi_1\phi_1}({\bf
k_1,k_2,k_3,-k_1-k_2-k_3},\omega_1,\omega_2,\omega_3,
-\omega_1-\omega_2-\omega_3) \nonumber \\
&=& \Gamma_{\hat\phi_2\phi_2\phi_2\phi_2}
({\bf k_1,k_2,k_3,-k_1-k_2-k_3},\omega_1,\omega_2,\omega_3,
-\omega_1-\omega_2-\omega_3) \nonumber \\
&=& 2\Gamma_{\hat\phi_1\phi_1\phi_1\phi_2}({\bf
k_1,k_2,k_3,-k_1-k_2-k_3},\omega_1,\omega_2,\omega_3,
-\omega_1-\omega_2-\omega_3) \nonumber \\
&=& 2\Gamma_{\hat\phi_2\phi_2\phi_2\phi_1}({\bf
k_1,k_2,k_3,-k_1-k_2-k_3},\omega_1,\omega_2,\omega_3,
-\omega_1-\omega_2-\omega_3) \nonumber \\
&=& \Gamma_{13}({\bf
k_1,k_2,k_3,-k_1-k_2-k_3},\omega_1,\omega_2,\omega_3,
-\omega_1-\omega_2-\omega_3)
\eea

Thus in the present model, the only
UV-divergent two- and four-point vertex functions which require
multiplicative renormalization are (i) $\partial_\omega
\Gamma_{11}({\bf k},\omega)$, (ii) $\partial_{k^2}\Gamma_{11}({\bf
k},\omega)$, (iii) $\Gamma_{11}({\bf k},\omega)$, (iv)
$\Gamma_{20}({\bf k},\omega)$ and (v) $\Gamma_{13}({\bf
k_1,k_2,k_3,-k_1-k_2-k_3},\omega_1,\omega_2,\omega_3,
-\omega_1-\omega_2-\omega_3)$.
 Each of them is to be rendered finite
through multiplicative renormalization by means of introducing a
$Z$-factor. Thus there are 5 $Z$-factors in total. However, there
are four parameters ($\Gamma,\,D,\,\tau,\,u$ and two fields
($\hat\phi_i,\,\phi_i,\,i=1\,or,\,2$), thus six altogether,
available for renormalization; thus this leaves us at liberty to
choose one of the renormalization constants in a convenient manner.

The renormalized kinetic coefficient $\Gamma_R$,
noise strength $D_R$, mass $\tau_R$ and coupling constant $u_R$ are defined in terms of the above vertex functions as
\begin{eqnarray}
&&\partial_\omega\Gamma_{11}(0,0)\equiv \frac{i}{\Gamma_R},\nonumber\\
&&\partial_{k^2}\Gamma_{11}(0,0)\equiv 1,\,\Gamma_{11}(0,0)\equiv
\tau_R,\,\Gamma_{20}(0,0) \equiv -\frac{2D_R}{\Gamma_R},\,
\Gamma_{13}({\bf k_i}=0,\omega_i=0)\equiv u_R.
\end{eqnarray}
The above definitions of the renormalized parameters allow us to calculate different renormalization $Z$-factors in the problem.

The perturbation theory here is constructed out of the bare
propagator and correlation functions which are to be read off from
the harmonic part of the action functional. From the generating
functional we get the bare propagators as
 \bea \langle
\phi_1({\bf k},\omega)\hat{\phi}_1(-{\bf k},-\omega)\rangle_0 &\equiv&
G_1^0(k,\omega)={i \over -{i \omega \over \Gamma} + \tau + k^2}=\langle \phi_2 ({\bf k},
\omega)\hat\phi_2(-{\bf k},-\omega)\rangle=G_2^0 (k,\omega) \nonumber \\
\eea and bare correlators as
 \bea
\langle \phi_1({\bf k},\omega)\phi_1(-{\bf k},-\omega)\rangle_0
&\equiv& C_1^0(k,\omega)= {2D \over \Gamma[{\omega^2 \over
\Gamma^2}+(\tau + k^2)^2]}=
\langle \phi_2({\bf k},\omega)\phi_2(-{\bf k},-\omega)\rangle_0 \equiv C_2^0(k,\omega), \nonumber \\
\langle \phi_1(k,\omega)\phi_2(-k,-\omega)\rangle_0 &\equiv&
C_x^0(k,\omega)={2i \hat{D}(k) \over \Gamma[{\omega^2 \over
\Gamma^2}+(\tau + k^2)^2]}. \eea

The self energy $\Sigma_G({\bf k},\omega)$ is formally given by the
Dyson equation:
 \bea
G_1^{-1}(k,\omega) = -{i\omega \over \Gamma} + \tau + k^2 -
\Sigma_G(k,\omega) = G_2^{-1}(k,\omega)=\Gamma_{11}(k,\omega).
 \eea
 In the same way one may define $\Sigma_D(k,\omega)$ through the
 relation
 \beq
 \Gamma_{20}(k,\omega)=2D +\Sigma_D(k,\omega).
 \eeq

 We now calculate fluctuation corrections to the relevant vertex functions. One-loop diagrammatic contributions
 to $\Gamma_{13}$ do not receive any contribution from $D_\times$ and are structurally identical to their equilibrium
 counterparts. Similarly one-loop corrections to $\Sigma_G({\bf
 k},\omega)$ are independent of $D_\times$ and are identical to
 the corresponding equilibrium contributions. In contrast, there are
 additional two-loop $D_\times$-dependent diagrammatic corrections
 to $\Sigma_G(k,\omega)$ and $\Sigma_D(k,\omega)$ whose evaluations require careful consideration.
 \begin{figure}[htb]
 \includegraphics[height=3.0cm,angle=0]{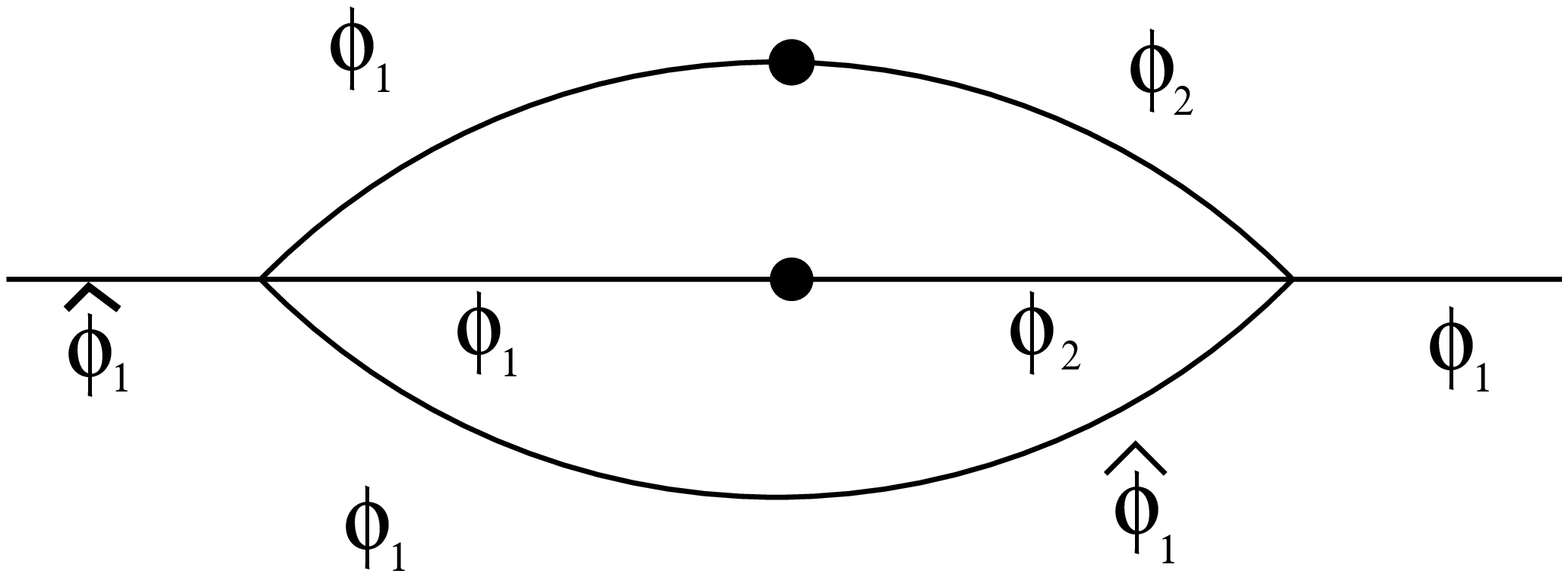} \hfill
\includegraphics[height=3.0cm,angle=0]{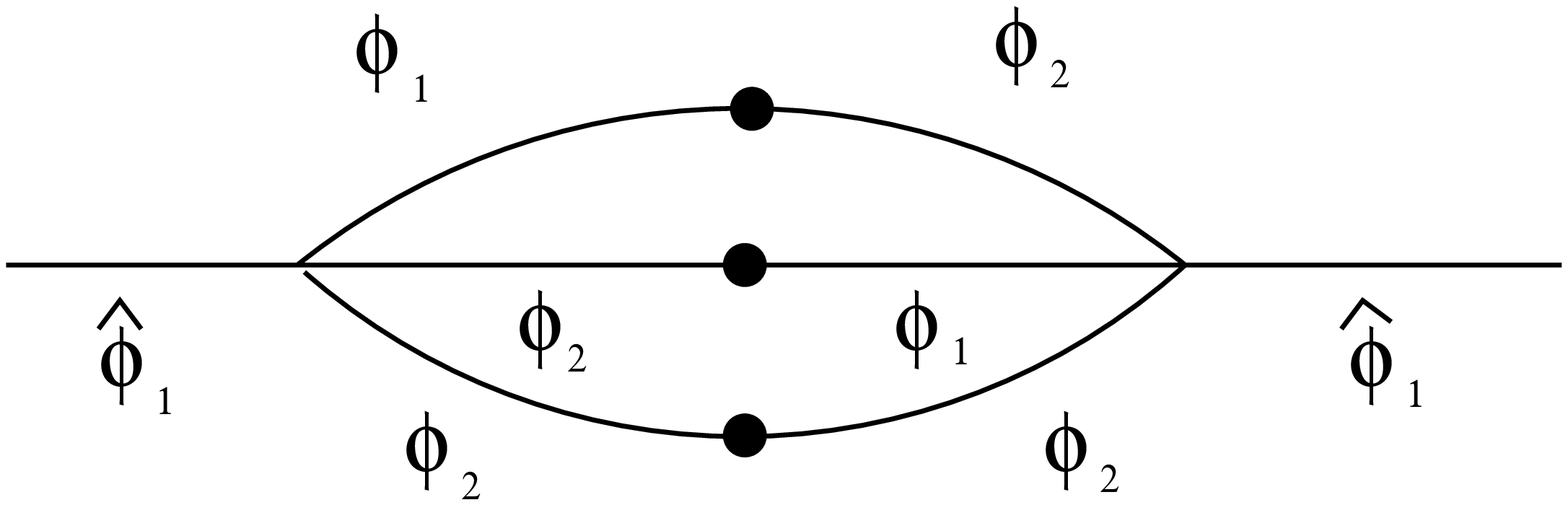}
\caption{Representative two-loop diagrams coming from non zero
noise cross correlations contributing to $\Sigma_G
(k,\omega)$ (left) and $\Sigma_D(k,\omega)$ (right). A line with a
filled circle represents a correlation function, a line without any
filled circle represents a propagator. We do not show all the
diagrams here.} \label{phase}
\end{figure}

 We
 separately consider the
 following contributions $\partial_\omega \Sigma_G
 (0,0)$, $\partial_{k^2}\Sigma_G(0,0)$ and $\Sigma_D(0,0)$. The cross-correlations contributions $\Sigma_G^\times (k,\omega)$
 and $\Sigma_D^\times (k,\omega)$ to $\Sigma_G(k,\omega)$ and $\Sigma_D(k,\omega)$ respectively, which
 do not arise in equilibrium, are all even in $\hat D({\bf q})$ (or in $D_\times$), since the model
 must be invariant under $D_\times \leftrightarrow -D_\times$. Such
 contributions to $\Sigma_G({\bf k},\omega)$ are of the form (up to
 numerical factor):
\begin{equation}
\Sigma_G^\times (k,\omega) =u^2\left({2 \over 6} - {1 \over 9}\right)
\int {d^dq_1 \over (2\pi)^d} {d^dq_2
\over (2\pi)^d} {{\hat D}({\bf q}_1) \over (\tau+q_1^2)}{{\hat
D}({\bf q}_2) \over (\tau+q_2^2)} {\Gamma \over [-i\omega +\Gamma
q_1^2 +\Gamma q_2^2+\Gamma \{3\tau+({\bf
k-q_1-q_2})^2\}]},\label{int1}
\end{equation}
and
\bea
\Sigma_D^\times (k=0,\omega=0) &=& u^2\left({1 \over 3} - {1 \over 18}\right)
\frac{1}{\Gamma}\int {d^dq_1 \over
(2\pi)^d} {d^dq_2 \over (2\pi)^d} {{\hat D}({\bf q}_1) \over
(\tau+q_1^2)}{{\hat D}({\bf q}_2) \over (\tau+q_2^2)} \frac{D}{\tau
+ ({\bf q}_1 + {\bf q}_2)^2}\nonumber \\
&&\frac{1}{3\tau + q_1^2 + q_2^2 + ({\bf
q}_1 + {\bf q}_2)^2}.\label{int2}
\eea
 where $\bf k$ and $\omega$ are external wavevector and
frequencies, respectively. We separately need to find out the
$k^0\omega$ and $k^2\omega^0$ parts of of the integral (\ref{int1}).
We need to calculate $\frac{\partial}{\partial a}\Sigma_G^\times
(k,\omega)|_{k=0,\omega=0}$ where $a=k^2,\,\omega$. Let us consider
$a=\omega$:
\begin{equation}
\frac{\partial}{\partial \omega} \int {d^dq_1 \over (2\pi)^d}
{d^dq_2 \over (2\pi)^d} {{\hat D}({\bf q}_1) \over
(\tau+q_1^2)}{{\hat D}({\bf q}_2) \over (\tau+q_2^2)} {1\over
-i\omega+\Gamma[3\tau+ q_1^2 + q_2^2+ ({\bf q}_1+{\bf
q}_2)^2]}.\label{intA}
\end{equation}
Since we are using an $\epsilon$-expansion based on a minimal
subtraction scheme, we need to extract the diverging parts of
(\ref{intA}), to be given by poles in $\epsilon$. It is noteworthy
that  integral (\ref{intA}) has a structure very similar to and has
the same logarithmic divergence (by simple power counting) as its
equilibrium counterpart [i.e., when $\hat D({\bf q})$ is replaced by
$D$ in (\ref{intA})]. Clearly, the dominant contribution to it comes
from ${\bf p}={\bf q}_1 + {\bf q}_2\sim 0$, which controls the
critical behavior of this integral. We write (retaining only the
small-${\bf p}$ contribution), up to constants and numerical factors
\begin{eqnarray}
\frac{\partial}{\partial \omega} \Sigma_G^\times
(k,\omega)|_{k=0,\omega=0} &\sim&\frac{\partial}{\partial\omega}\int
{d^dq_1 \over (2\pi)^d} {d^dq_2 \over (2\pi)^d} {{\hat D}({\bf q}_1)
\over (\tau+q_1^2)} {{\hat D}({\bf p}-{\bf q}_1) \over
(\tau+q_2^2)} {1\over -i\omega+\Gamma[3\tau+ q_1^2 + q_2^2+ p^2]}\nonumber \\
&\sim& \frac{\partial}{\partial\omega}\int {d^dq_1 \over (2\pi)^d}
{d^dq_2 \over (2\pi)^d} {{\hat D}({\bf q}_1) \over (\tau+q_1^2)}
{{\hat D}({\bf q}_1) \over (\tau+q_2^2)} {1\over
-i\omega+\Gamma[3\tau+ q_1^2 + q_2^2+ p^2]}\nonumber \\
&=&D_\times^2\frac{\partial}{\partial \omega} \int {d^dq_1 \over
(2\pi)^d} {d^dq_2 \over (2\pi)^d} {1 \over (\tau+q_1^2)}{1 \over
(\tau+q_2^2)} {1\over {-i\omega+\Gamma[3\tau+ q_1^2 + q_2^2+ ({\bf
q}_1+{\bf q}_2)^2]}}\nonumber \\ &=&-i\frac{D_\times^2}{3\Gamma^2} 
\int {d^dq_1 \over (2\pi)^d} {d^dq_2 \over (2\pi)^d} {1 \over
(\tau+q_1^2)}{1 \over (\tau+q_2^2)}\frac{1}{\tau + ({\bf q}_1 + {\bf
q}_2)^2} \frac{1}{3\tau + q_1^2 + q_2^2 + ({\bf q}_1 +{\bf q}_2)^2},\nonumber \\
\label{approx1}
\end{eqnarray}
where, we have replaced integral (\ref{intA}) by its dominant
contribution coming from ${\bf p}\sim 0$. The last line of
Eq.(\ref{approx1}) is obtained by symmetrizing the previous line.
The reduced integral (\ref{approx1}) may now be evaluated exactly in
the same way just as its equilibrium counterpart (i.e., when
$D_\times$ is replaced by $D$). In a similar way, cross-correlation
contribution to the anomalous dimension (again logarithmically
divergent on a na\"ive power counting basis) may be written as
(up to constants and numerical factors)
\begin{equation}
\frac{\partial}{\partial k^2} \Sigma_G^\times(k,\omega)|
_{k=0,\omega=0}\sim D_\times^2 \frac{\partial}{\partial k^2}  \int
{d^dq_1 \over (2\pi)^d} {d^dq_2 \over (2\pi)^d} {1 \over
(\tau+q_1^2)}{1 \over (\tau+q_2^2)} \frac{1}{\tau + ({\bf k} + {\bf
q}_1 + {\bf q}_2)^2},\label{approx2}
\end{equation}
and the cross-correlation contributions to noise strengths
(\ref{int2}) become (up to constants and numerical factors)
\begin{equation}
\Sigma_D^\times (0,0)\sim DD_\times^2   \int {d^dq_1 \over (2\pi)^d} {d^dq_2
\over (2\pi)^d} {1 \over (\tau+q_1^2)}{1 \over (\tau+q_2^2)}
\frac{1}{\tau + ( {\bf q}_1 + {\bf q}_2)^2} \frac{1}{3\tau + q_1^2 +
q_2^2 + ({\bf q}_1 + {\bf q}_2)^2}.\label{approx3}
\end{equation}
In the above, in our evaluations of the cross-correlation
contributions $\frac{\partial}{\partial a}\Sigma_G^\times
(k,\omega)_{k=0,\omega=0},\,a=k^2,\omega$ and $\Sigma_D^\times
(0,0)$, we have picked up the dominant contribution given by ${\bf
p}={\bf q}_1 + {\bf q}_2\sim 0$. Subdominant contributions are
neglected and are expected to be {\em small} as we heuristically
justify: For example, for ${\bf p}\gg {\bf q}_1$, the integrand in
$\Sigma^\times_D (0,0)$ (see Eq.~(\ref{approx3} above) is
\begin{eqnarray}
&&{\hat D({\bf q}_1) \over (\tau+q_1^2)}{\hat D({\bf q}_2) \over
(\tau+q_2^2)}\frac{D} {\tau + ({\bf q}_1 + {\bf q}_2)^2} \frac{1}
{3\tau + q_1^2 + q_2^2 + ({\bf q}_1 + {\bf q}_2)^2} \sim {\hat
D({\bf q}_1) \over (\tau+q_1^2)}\frac{\hat D({\bf p})}{\tau + p^2}
\frac{D}{\tau + p^2}\frac{1}{3\tau + 2p^2}, \nonumber \\
\end{eqnarray}
which can be both +ve and -ve since $\hat D({\bf q}_1)$ and $\hat
D({\bf p})$ are odd functions of their arguments, and hence
contributions from outside the dominant region ${\bf p}\sim 0$ will
be small due to mutual cancelations. Our results, although backed up
by heuristic arguments, nevertheless bring out remarkable new
features, as we shall see below. Thus, after putting every
diagrammatic contribution (up to two-loop order) together we obtain
for $\Sigma_G({\bf k},\omega)$ as
 \bea
\Sigma_G(k,\omega)&=& \frac{2uD}{3}\int \frac{d^dq}{(2\pi)^2} \frac{1}{\tau+q^2} + \Gamma u^2D^2\left( {1 \over 2} + {1 \over 18} + {1 \over 9}\right)
\int\frac {d^2q_1}{(2\pi)^d}\frac {d^dq_2}{(2\pi)^d} {1 \over \tau+q_1^2} \nonumber \\
&\times & {1 \over \tau+q_2^2}{1 \over {-i\omega +
\Gamma\{3\tau+q_1^2+q_2^2+({\bf k}-{\bf q}_1-{\bf q}_2)^2\}}}
 \nonumber \\ &+& \Gamma u^2 D_\times^2\left( {2 \over 6} - {1 \over 9}\right)\int\frac {d^2q_1}{(2\pi)^d}\frac {d^dq_2}{(2\pi)^d}
 {1 \over \tau+q_1^2} {1 \over \tau+q_2^2}{1
\over {-i\omega + \Gamma\{3\tau+q_1^2+q_2^2+({\bf k}-{\bf q}_1-{\bf
q}_2)^2\}}}. \nonumber \\ \eea
 Similarly the two loop contributions to
$\Sigma_D(0,0)$ comes out to be
 \bea
\Sigma_D(k=0,\omega=0) &=& {u^2D^3 \over \Gamma}\left( {1 \over 6} + {1 \over 18} \right)\int \frac {d^2q_1}{(2\pi)^d}\frac {d^dq_2}{(2\pi)^d}
 {1 \over \tau+q_1^2} {1 \over \tau+q_2^2} \frac{1}{3\tau+q_1^2+q_2^2 + ({\bf q}_1+{\bf q}_2)^2} \nonumber \\
&& {1 \over \tau+({\bf q}_1+{\bf q}_2)^2} +{u^2DD_\times^2 \over \Gamma}\left({1 \over 3}- {1 \over 18}\right)\int \frac {d^2q_1}{(2\pi)^d}\frac {d^dq_2}{(2\pi)^d}
{1 \over \tau +q_1^2} {1 \over \tau +q_2^2}\nonumber \\
&\times&{1 \over \tau+({\bf q}_1+{\bf q}_2)^2}\frac{1}{3\tau+q_1^2+q_2^2 + ({\bf q}_1+{\bf q}_2)^2}.
 \eea
 Although formally there exists
two-loop diagrammatic corrections (there are no one-loop
corrections) to $\hat D ({\bf k})$, all of these vanish in the long
wavelength limit due to the fact that $\hat D ({\bf k})$ is an odd
function of wavevector $\bf k$. In Fig.~(\ref{cross}) we consider
one such two-loop diagram:
\begin{figure}[htb]
 \includegraphics[height=3.0cm,angle=0]{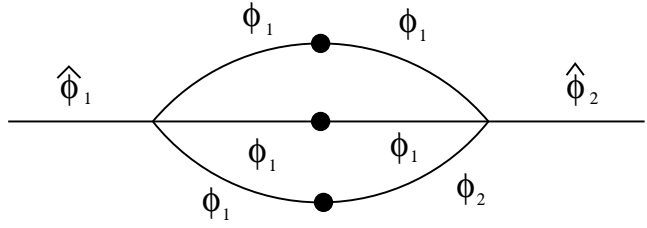}
\caption{A representative two-loop diagram contributing to
$\Sigma_\times (k,\omega)$. Symbols have meanings as before.}
\label{cross}
\end{figure}

The corresponding expression $\Sigma_\times (k,\omega)$ is (up to
constants and numerical factors)
\begin{eqnarray}
\Sigma_\times (k,\omega) \sim \int\frac{d^dq_1}{(2\pi)^d}
\frac{d^dq_2}{(2\pi)^d} {\hat D ({\bf q}_1) \over \tau+q_1^2} {1
\over \tau+q_2^2} {1 \over \tau+({\bf q}_1+{\bf
q}_2)^2}\frac{1}{3\tau+q_1^2+q_2^2 + ({\bf q}_1+{\bf q}_2)^2}.
\label{sigmacross}
\end{eqnarray}
Clearly, the integral in (\ref{sigmacross}) vanishes. We shall come
back to this issue of non-renormalization of $D_\times ({\bf k})$
again at the end.

Finally, to complete evaluating diagrammatic corrections we now evaluate
the $\Gamma_{13}$ up to one-loop order at zero external wavevector and frequency. There are no
contributions from $D_\times$ to the four point vertex function. We
obtain
 \bea \Gamma_{(3,1)}=u^2D{\mu^{-\epsilon} \over
2(4\pi)^d}\Gamma(\epsilon/2)\left[{1 \over 4}+ {1 \over 36}\right].
\eea

After evaluating all the two point and four point vertex functions
we can now renormalize vertex functions $\Gamma_{(11)}(0,0)$,
${\partial \over
\partial \omega}\Gamma_{(11)}(\omega=0,0)$, ${\partial \over
\partial k^2} \Gamma_{(11)}(0,k=0)$ and $\Gamma_{(3,1)}(0,0)$. We now define renormalization
$Z$-factors for the present model. We begin by introducing the
renormalized fields $\phi_i^R$ and $\hat\phi^R_i$:
\begin{equation}
\phi^R_i=Z^{1/2}\phi_i,\;\;\;\hat{\phi}^R_i=\hat{Z}^{1/2}\hat{\phi}_i.
\end{equation}
This implies that the renormalized vertex functions become
 \bea
\Gamma^R_{(N,\hat{N})}=Z^{-N/2}\hat{Z}^{-\hat{N}/2}\Gamma_{(N,\hat{N})}.
 \eea
 We further define the renormalized parameters as
  \bea D^R=Z_DD
\, , \,\, \tau^R=Z_\tau \tau \mu^{-2} \,,\,\, u^R=Z_uu A_d
\mu^{d-4}, \, , \,\, \Gamma^R=Z_\Gamma \Gamma,
 \eea
  where $\mu$ is the scale
factor introduced to make the renormalized parameters dimensionless.
Here $A_d={1 \over 2^{d-1}\pi^{d \over 2}}$. Thus we get
 \bea
{\partial \over \partial \omega}\Gamma^R_{(11)}(\omega=0,0) &=& Z^{-1/2}\hat{Z}^{-1/2}{i \over \Gamma}
\left[1+ {2D^2u^2\mu^{-2\epsilon} \over 3(4\pi)^d\epsilon}\ln({4 \over 3}) +
 {2 D_\times^2u^2\mu^{-2\epsilon} \over 9(4\pi)^d\epsilon}\ln({4 \over 3})\right]\nonumber \\
&\equiv& {i \over \Gamma^R}={i \over Z_\Gamma \Gamma} \\
{\partial \over \partial k^2}\Gamma^R_{(11)}(0,k=0) &=& iZ^{-1/2}\hat{Z}^{-1/2}
\left[1+{D^2u^2\mu^{-2\epsilon} \over 9(4\pi)^d \epsilon} + {D_\times^2u^2\mu^{-2\epsilon} \over 27(4\pi)^d\epsilon}\right] \nonumber \\
&=& i \\
\Gamma^R_{(20)}(0,0) &=& -\hat{Z}^{-1}{2D \over \Gamma}
\left[1+ {2D^2u^2\mu^{-2\epsilon} \over 3(4\pi)^d\epsilon}\ln({4 \over 3}) + {5 D_\times^2u^2\mu^{-2\epsilon} \over 6(4\pi)^d\epsilon}\ln({4 \over 3})\right] \nonumber \\
&=& -{2D^R \over \Gamma^R} \\
\Gamma^R_{(3,1)}(0,0) &=& \hat{Z}^{-1/2}Z^{-3/2}{u \over 6}\left[ 1 - {10 \over 6\epsilon}Du\mu^{-\epsilon}\right] \\
\Gamma^R_{(11)}(0,0) &=& Z^{-1/2}\hat{Z}^{-1/2} \tau \left[1- {4uD\mu^{-\epsilon} \over 3(4\pi)^{d/2}\epsilon} -
{2D^2u^2\mu^{-2\epsilon} \over 3(4\pi)^d\epsilon}\left( {2 \over \epsilon} +1\right) -
{2 D_\times^2u^2\mu^{-2\epsilon} \over 9(4\pi)^d\epsilon}\left( {2 \over \epsilon}+1\right) \right] \nonumber \\
&=& \tau^R,
 \eea
  from which we can calculate all the $Z$-factors. We use the freedom to
  choose one of the $Z$-factors freely to set $Z_D=1$. Henceforth we set
$D=1$ for simplicity without any loss of generality.
Assuming $D_\times^2=N_\times D^2$, where $N_\times$ is any dimensionless
parameter, the other $Z$ factors are obtained up to two loop order
as follows
 \bea
Z_\Gamma &=& 1- {1 \over 36}{(uA_d\mu^{-\epsilon})^2 \over \epsilon}\left( 6\ln{4 \over 3} -1\right)
- {1 \over 108}N_\times {(uA_d\mu^{-\epsilon})^2 \over \epsilon}\left( 6\ln{4 \over 3} -1\right) \\
\hat{Z} &=& 1 + {1 \over 36}{(uA_d\mu^{-\epsilon})^2 \over \epsilon} +
{11 \over 72}N_\times {(uA_d\mu^{-\epsilon})^2 \over \epsilon}\ln{4 \over 3} + {1 \over 108} N_\times {(uA_d\mu^{-\epsilon})^2 \over \epsilon} \\
Z &=& 1 + {1 \over 36}{(uA_d\mu^{-\epsilon})^2 \over \epsilon} - {11 \over 72} N_\times {(uA_d\mu^{-\epsilon})^2 \over \epsilon}\ln{4 \over 3}
+ {1 \over 108} N_\times {(uA_d\mu^{-\epsilon})^2 \over \epsilon} \\
Z_\tau &=& 1 - {2 \over 3} {(uA_d \mu^{\epsilon}) \over \epsilon}
 - {1 \over 18} \left( {7 \over 2} + {6 \over \epsilon}\right){(uA_d\mu^{-\epsilon})^2 \over \epsilon}
 - {1 \over 54} N_\times\left( {7 \over 2} + {6 \over \epsilon}\right) {(uA_d\mu^{-\epsilon})^2 \over \epsilon}
 \eea

Defining the Wilson's flow functions as
 \bea
\zeta_\phi &=& \mu{\partial \over \partial \mu}\ln{Z} \, , \,\, \zeta_{\hat{\phi}}=\mu{\partial \over \partial \mu}\ln{\hat{Z}} , \\
\zeta_\Gamma&=&\mu{\partial \over \partial \mu}\ln{Z_\Gamma}, \\
\zeta_D &=& \mu{\partial \over \partial \mu}\ln Z_D, \\
\zeta_\tau &=& \mu{\partial \over \partial \mu}\ln{\tau^R \over
\tau} = -2 + \mu{\partial \over \partial \mu}\ln Z_\tau,
 \eea and
the $\beta$ function for the non-linear coupling as
 \bea
\beta_u=\mu{\partial \over \partial \mu}u^R=u\left(-\epsilon + {10
\over 6}u\right)
 \eea
  we get a stable nontrivial fixed point at $u^*={6 \over 10}\epsilon$ and
we can evaluate the critical exponents from these flow functions at
the fixed point. The flow functions at the fixed point pick up
values up to order $\epsilon^2$ as follows:
 \bea
\zeta_{\phi} &=&-{1 \over 50}\epsilon^2 - {1 \over 150}N_\times\epsilon^2 + {11 \over 100}N_\times\epsilon^2\ln{4 \over 3} \label{zetaphi} \\
\zeta_{\hat{\phi}}&=& -{1 \over 50}\epsilon^2 - {1 \over 150}
N_\times\epsilon^2
 - {11 \over 100}N_\times\epsilon^2\ln{4 \over 3} \\
\zeta_\Gamma &=& {1 \over 50}\epsilon^2\left(6\ln{4 \over 3} -1 \right) +
 {1 \over 150}N_\times\epsilon^2\left( 6\ln{4 \over 3} - 1\right) \label{zetagamma}\\
\zeta_\tau &=& -2 + {2 \over 5}\epsilon + O(\epsilon^2).\label{zetatau}
 \eea

The basic renormalization group(RG) equation is derived on the basis that the unrenormalized correlation
 and vertex functions do not depend on the arbitrary scale $\mu$. Hence if we hold the bare parameters $D, \tau$ and $\mu$ fixed, we must have
 \bea
0 &=&\mu{d \over d\mu}|_{D,\tau,\mu}\Gamma_{N,\hat{N}} \nonumber \\
&=& \mu{d \over d\mu}[\hat{Z}^{\hat{N}/2}Z^{N/2}\Gamma^R_{N,\hat{N}}(\mu,\Gamma^R,\tau^R,u^R)].
 \eea
As $Z$-factors also depend on $\mu$, the RG equation finally takes
the form
 \bea \hat{Z}^{\hat{N}/2}Z^{N/2}\mu\left[ {\partial \over
\partial \mu} + {\hat{N}/2}{\partial \over \partial \mu}\ln\hat{Z} +
{N \over 2}{\partial \over \partial \mu}\ln{Z} + {\partial\Gamma^R
\over \partial \mu}{\partial \over \partial\Gamma^R} + {\partial
\tau^R \over \partial \mu}{\partial \over \partial\tau^R} +
{\partial u^R \over \partial \mu}{\partial \over \partial
u^R}\right]\Gamma^R_{N,\hat{N}}=0.\nonumber \\
 \eea

At the critical point we have scale invariance separately under
scaling of space, time fields and parameters. These are determined
by the momentum and frequency canonical dimensions of the fields and
parameters. After proper scaling as described above in this Section
we have canonical dimensions of fields and parameters as
\bea
d^k_{\phi_1,\phi_2}=d/2-1 \,\, &,& \,\,d^\omega_{\phi_1,\phi_2}=0, \nonumber \\
d^k_{\hat{\phi}_1,\hat{\phi}_2}=d/2-1 \,\, &,& \,\, d^\omega_{\hat{\phi}_1,\hat{\phi}_2}=1 \nonumber \\
d^k_\Gamma = -2 \,\, &,& \,\, d^\omega_\Gamma=1 \nonumber \\
d^k_\tau =0 \,\, &,& \,\, d^\omega_\tau=0. \eea

Canonical scale invariance at the fixed point ($\beta_u^*=0$) for
the correlation function implies
 \bea
\left[\mu{\partial \over \partial \mu} - x{\partial \over \partial x} - 2\Gamma^R{\partial \over \partial \Gamma^R}
  - d^k_C\right]C^R(x,t,\mu,\Gamma^R,\tau^R) &=& 0 \,\,\,\,\, \mbox{and} \nonumber \\
\left[ \Gamma^R {\partial \over \partial \Gamma^R} - t{\partial
\over \partial t} - d^\omega_C\right]C^R(x,t,\mu,\Gamma^R,\tau^R)
&=& 0 \label{tot}
 \eea
where $d^k_C$ and $d^\omega_C$ are the momentum and frequency
dimension of the correlation function $C(x,t)$ respectively. In this
case $d_C^k=d-2$ and $d^\omega_C=0$. Therefore from Eq.~(\ref{tot})
we have \bea \Gamma^R{\partial \over \partial
\Gamma^R}C^R(bx,b^2t,\mu,\Gamma^R,\tau^R)=t{\partial \over \partial
t}C^R(x,t,\mu,\Gamma^R,\tau^R) \label{temp} \eea The RG equation for
the correlation function at the fixed point can be written as \bea
0 &=& \mu{\partial \over \partial\mu}C = \mu{\partial \over \partial\mu}[Z^{-1}C^R] \nonumber \\
&=& \left[ \mu{\partial \over \partial \mu} - \zeta_\phi +
\zeta_\Gamma\Gamma^R{\partial \over \partial\Gamma^R} +
\zeta_\tau\tau^R{\partial \over \partial \tau^R}\right]C^R(x, t,\mu,
\Gamma^R,\tau^R) \label{RG} \eea

Combining the two separate spacial and temporal scale invariant
equations Eqs.~(\ref{tot}) and using equations (\ref{temp}) and
(\ref{RG}) we get at the fixed point
 \bea
  \left[-x{\partial \over
\partial x} -\zeta_\tau\tau^R{\partial \over \partial \tau^R} -
(2+\zeta_\Gamma)t{\partial \over \partial t} +
{2-d-\eta}\right]C^R(x,t,\Gamma^R,\tau^R)=0, \label{final}
 \eea
where we have used $\eta=-\zeta_\phi$. From Eq.~(\ref{final}) it can
be seen that at the critical point $(\tau=0)$ and equal time $(t=0)$
the correlation function should take the form
 \bea C(x)\sim
x^{2-d-\eta},
 \label{eta} \eea
  which gives the spatial scaling of the equal-time correlation function at the critical point.
In case of time-dependent correlation function at the critical point
the scale invariant equation takes the form
 \bea \left[-x{\partial
\over \partial x} - (2+\zeta_\Gamma)t{\partial \over \partial t} +
2-d-\eta\right]C^R(x,t,\Gamma^R)=0. \eea
 Assuming dynamical
scaling, the solution of $C(x,t)$ should be of the form $C(x,t)\sim
x^{2-d-\eta}g(x^{2+\zeta_\Gamma}/t)$, which implies that
 \bea
2+\zeta_\Gamma=z \label{dexp}
 \eea
  should be the dynamic exponent.
At equal time $(t=0)$, near the critical point $(\tau^R\neq 0)$, the
equation for $C(x,t)$ can be written as
 \bea
  \left[-x{\partial \over
\partial x} -\zeta_\tau\tau^R{\partial \over \partial \tau^R} +
2-d-\eta\right]C^R(x,t,\tau^R)=0. \eea
 This implies that the
correlation function should be of the form $C(x,\tau)\sim
x^{2-d-\eta}f(x^{\zeta_\tau}/\tau)$, and the correlation length
exponent
 \bea \nu=-{1 \over \zeta_\tau}. \label{nu}
  \eea

For the propagator $G=\langle\phi\hat{\phi}\rangle$, the scale invariant equations are given by
\bea
\left[\mu{\partial \over \partial \mu} - x{\partial \over \partial x} - 2\Gamma^R{\partial \over \partial \Gamma^R}
- d^k_G\right]G^R(x,t,\mu,\Gamma^R,\tau^R) &=& 0 \,\,\,\,\, \mbox{and} \nonumber \\
\left[ \Gamma^R {\partial \over \partial \Gamma^R} - t{\partial
\over \partial t} - d^\omega_G\right]G^R(x,t,\mu,\Gamma^R,\tau^R)
&=& 0. \label{propscale} \eea where $d^k_G=d-2$ and $d^\omega_G=1$.
From the second of equations (\ref{propscale}) it is obvious that
\bea \Gamma^R{\partial \over \partial
\Gamma^R}G^R(x,t,\mu,\Gamma^R,\tau^R)=\left[1+t{\partial \over
\partial t}\right]G^R(x,t,\mu,\Gamma^R,\tau^R) \label{tempprop}
\eea

The RG equation for the propagator at the fixed point can be written as
\bea
0 &=& \mu{\partial \over \partial\mu}G = \mu{\partial \over \partial\mu}[Z^{-1/2}\hat{Z}^{-1/2}G^R] \nonumber \\
&=& \left[ \mu{\partial \over \partial \mu} - {1 \over 2}\zeta_\phi
- {1 \over 2}\zeta_{\hat{\phi}} + \zeta_\Gamma\Gamma^R{\partial
\over \partial\Gamma^R} + \zeta_\tau\tau^R{\partial \over \partial
\tau^R}\right]G^R(x,t,\mu, \Gamma^R,\tau^R) \label{RGprop} \eea

Using equations (\ref{RGprop}) and (\ref{tempprop}) in (\ref{propscale}) we get
\bea
 \left[-x{\partial \over
\partial x} -\zeta_\tau\tau^R{\partial \over \partial \tau^R} -
(2+\zeta_\Gamma^R)\{1+t{\partial \over \partial t}\} + {2-d-{1 \over
2}\eta - {1 \over 2}\hat{\eta}}\right]G^R(x,t,\Gamma^R,\tau^R)=0,\nonumber \\
\label{finalprop} \eea where $\hat{\eta}=-\zeta_{\hat{\phi}}$.

From Eq.~(\ref{finalprop}) the time dependent propagator at the
critical point $(\tau=0)$ can be written as \bea \left[-x{\partial
\over \partial x} - (2+\zeta_\Gamma)\{1+t{\partial \over \partial
t}\} + 2-d-{1 \over 2}\eta - {1 \over 2}\hat{\eta}
\right]G^R(x,t,\Gamma^R)=0. \eea

Assuming dynamical scaling $G(x,t)$ should be of the form $G\sim
x^{2-d-\eta/2 - \hat{\eta}/2 - 2-\zeta_\Gamma} f({x^{2+\zeta_\Gamma}
\over t})$. Since the static susceptibility $\chi(x)$ is
proportional to $\int_0^\infty dt G(x,t)$ which on integrating over
time gives us
 \bea
 \chi(x)\sim x^{2-d-\eta/2-\hat{\eta}/2}.
 \eea

From equations (\ref{dexp}), (\ref{eta}) and (\ref{nu}) and using
equations (\ref{zetagamma}), (\ref{zetaphi}) and (\ref{zetatau}) we
get the dynamic exponent, anamolous dimension and correlation length
exponent of the model to the leading order in $\epsilon$:
 \bea
z &=& 2+{1 \over 50}\epsilon^2\left(6\ln{4 \over 3} - 1\right) + {1 \over 150}N_\times\epsilon^2\left(6\ln{4 \over 3} - 1\right) \\
\eta &=& {1 \over 50}\epsilon^2 + {1 \over 150}N_\times\epsilon^2 - {11 \over 100}N_\times\epsilon^2\ln{4 \over 3},
 \\
 \hat \eta &=& {1 \over 50}\epsilon^2 + {1 \over 150}N_\times\epsilon^2 + {11 \over 100}N_\times\epsilon^2\ln{4 \over
 3},\\
{1 \over \nu} &=& 2-{2 \over 5}\epsilon + O(\epsilon^2).\eea
 If $N_\times=0$ we get back the equilibrium
exponents as expected. Their contribution from the nonequilibrium
part depends on the value of $N_\times$. Let us now consider the
consequences of a non-zero $N_\times$. First of all, as is evident
from the results presented above, the static susceptibility and the
equal-time correlation function {\em do not} display the same
spatial scaling, since $\eta\neq\hat\eta$. This is an important
evidence of breakdown of the FDT in the renormalized theory. Note,
in the linearized theory  breakdown of the FDT is manifested in the
existence of non-zero cross-correlation functions: The correlation
matrix pick up non-zero off-diagonal elements, where as the dynamic
susceptibility matrix remains diagonal. Nevertheless, all elements
of the correlation and the dynamic susceptibility matrix exhibit the
same scaling properties at the critical point. In contrast, in the
renormalized non-linear theory, not only the correlation matrix is
off-diagonal where as the susceptibility matrix remains diagonal,
the elements of the correlation matrix scale differently from those
of the susceptibility matrix. The latter result is purely a
non-linear effect. Further, the dynamic exponent for finite
$N_\times$, $z(N_\times)$ is larger than $z(N_\times =0)$, its value
for equilibrium dynamics. Thus relaxation for finite $N_\times$ is
{\em slower} than for the corresponding equilibrium dynamics. The
correlation length exponent $\nu$ has been calculated only up to
$O(\epsilon)$ and is equal to its equilibrium value. However, as
there are $N_\times$-dependent corrections to $\Sigma_G (0,0)$ at
the two-loop order, the value of $\nu$ is likely to be different
from its equilibrium value at the two-loop order. Lastly, the
$N_\times$-dependence of all the scaling exponents are {\em
continuous} - all of them vary continuously with $N_\times$ and go
over to their equilibrium values when $N_\times$ is set to zero. Our
claim of the scaling exponents varying continuously with $N_\times$
rests on the marginality of $N_\times$. We have shown explicitly up
to two-loop order that $\hat D ({\bf k})$ does not renormalize.
Hence $N_\times$ does not renormalize up to two-loop order and
depends on the bare value of $D_\times^2$. Any non-zero fluctuation
corrections to $\hat D({\bf q})$ must be an odd function of the its
wavevector argument. In order to have that one must have odd number
of internal cross-correlation line. Since all internal wavevectors
are integrated over, such a contribution will vanish in the limit of
vanishing external wavevector and frequency. Thus $\hat D({\bf q})$
and remains unrenormalized and hence $N_\times$ appears as a
dimensionless marginal operator to any order in perturbation.

\section{Summary}
\label{conclu} To summarize, we have analyzed the universal scaling
properties of a nonequilibrium version of $O(2)$-symmetric dynamical
model near the critical point. We write down a non-conserved
relaxational dynamics for the order parameter field. We have
introduced cross-correlations between the two additive noises in the
Langevin equations, so that the FDT is immediately broken. We then
show that if the cross-correlation is imaginary and odd in
wavevector, the underlying $O(2)$ symmetry is still maintained. We
calculate the scaling exponents of the model in a DRG framework
using an $\epsilon$-expansion scheme, where $\epsilon=4-d$ with 4
being the upper critical dimension of the model. We show that at the
two-loop order there are diagrammatic corrections to the various
two-point vertex functions in the model arising from the
cross-correlations. We have used heuristic arguments to extract the
dominant contributions to the two-loop diagrams involving
cross-correlations, which have allowed us to evaluate the respective
cross-correlation contributions in a simple and controlled manner.
We finally argue that the cross-correlation amplitude appears as a
marginal operator in the problem. Since this amplitude appears in
the expressions of the scaling exponents we have an example of a
continuously varying universality class. Technically speaking we
obtain a {\em fixed line}, parametrized by the value of the
parameter $N_\times$ introduced above, instead of a single or
isolated fixed points. Every point on the fixed line characterizes a
universality class, parametrized again by $N_\times$. The fixed line
begins from $N_\times=0$, which is the equilibrium fixed point. This
stands in contrast to, e.g., Ref.~\cite{tauber-etal:02}, where
nonequilibrium noises lead to additional fixed points, but not a
fixed line as here. There are other dynamical models where
cross-correlated noises lead to universal properties varying
continuously with the amplitude of the noise cross-correlations.
Notable examples are the stochastically driven generalized Burgers
model (GBM) \cite{abfreylong} and magnetohydrodynamic turbulence
(MHD) \cite{abmhd}. However, the $d$-dimensional GBM and MHD models
are intrinsically nonequilibrium and do not generally have an
equilibrium limit: Switching off the noise cross-correlation does
not make these models equilibrium in general $d$-dimensions. In
contrast, the present model has a well-defined equilibrium limit
given by $N_\times=0$ for any dimension $d$. Thus not only does our
model here exhibit continuously varying universal properties, it can
be driven away from equilibrium continuously and incrementally by
tuning $N_\times$. Continuously varying universality has been
found in Ref.~\cite{uwe2} as well. However, Ref.~\cite{uwe2}
required coupling of the order parameter field with a conserved
density. In contrast in our work we have the order parameter field
only as the relevant dynamical field. Quantitative accuracy of our
results is limited by the heuristic arguments we resorted to while
evaluating the diagrams arising from noise cross-correlations. In
order to verify this, direct numerical simulations of the model
Langevin equations, or simulations of appropriately defined
lattice-gas models should be performed. In the present article we
have discussed the universal scaling properties at the critical
point only. Numerical simulations of a driven $O(3)$ model
\cite{sriram-jayajit} displays existence of spatio-temporal chaotic
low-temperature regime below its critical point in the absence of
stochastic noises. This chaos, when {\em controlled}, is replaced by
spatially periodic steady helical states which are robust against
noise. In view of these results in Ref.~\cite{sriram-jayajit}, it
would be interesting to examine the properties of the ordered phase
below $T_c$, and their dependences on the parameter $N_\times$
introduced above. In the above we have confined ourselves in
discussing a usual order-disorder transition and the associated
universality at the critical point. For any model, such a scenario
holds as long as the physical dimension is greater than the lower
critical dimension $d_L$ of the model. For equilibrium models with
continuous symmetries, e.g., the $O(2)$-symmetric model in
equilibrium, the Mermin-Wagner theorem tells us that $d_L=2$. For
models out of equilibrium, there are no such general results. It
would be interesting to examine the present model, perhaps through
non-perturbative methods, at $d=2$, in particular the role and
dynamics of topological defects in the presence of noise
cross-correlations. We hope our theoretical results will inspire
more detailed theoretical studies on more realistic models or
experimental work on driven systems with coupled variables, where
the role of noise cross-correlations in determining the universal
properties may be explicitly tested.

\section{Acknowledgement}
We thank J. K. Bhattacharjee for useful discussions and critical comments.
We also thank U. C. T\"auber and J. E. Santos for helpful suggestions. 
One of the authors (AB) wishes to thank the
Max-Planck-Society (Germany) and Department of Science and
Technology (India) for partial support through the Partner Group
programme (2009).


\end{document}